\documentclass[prd,twocolumn,superscriptaddress]{revtex4}
\usepackage{graphicx}
\begin{document}

\newcommand{\Msun}{\mbox{$\rm M_{\odot}$}}

\title{The onset of the non-linear regime in unified dark matter models}

\author{P.P. Avelino}
\email[Electronic address: ]{ppavelin@fc.up.pt}
\affiliation{Centro de F\'{\i}sica do Porto e Departamento de F\'{\i}sica da Faculdade de
Ci\^encias da Universidade do Porto, Rua do Campo Alegre 687,
4169-007, Porto, Portugal} 
\affiliation{Astronomy Centre, University of Sussex, 
Brighton BN1 9QJ, United Kingdom}
\author{L.M.G. Be\c ca}
\email[Electronic address: ]{luis.beca@fc.up.pt}
\affiliation{Centro de F\'{\i}sica do Porto e Departamento de F\'{\i}sica da Faculdade de
Ci\^encias da Universidade do Porto, Rua do Campo Alegre 687,
4169-007, Porto, Portugal}
\author{J.P.M. de Carvalho}
\email[Electronic address: ]{mauricio@astro.up.pt}
\affiliation{Centro de Astrof\'{\i}sica da Universidade do Porto,
R. das Estrelas s/n, 4150-762 Porto, Portugal}
\affiliation{Departamento de Matem\'atica Aplicada da Faculdade de
Ci\^encias da Universidade do Porto, Rua do Campo Alegre 687,
4169-007, Porto, Portugal}
\author{C.J.A.P. Martins}
\email[Electronic address: ]{C.J.A.P.Martins@damtp.cam.ac.uk}
\affiliation{Centro de Astrof\'{\i}sica da Universidade do Porto,
R. das Estrelas s/n, 4150-762 Porto, Portugal}
\affiliation{Department of Applied Mathematics and Theoretical
Physics, Centre for Mathematical Sciences,\\ University of
Cambridge, Wilberforce Road, Cambridge CB3 0WA, United Kingdom}
\affiliation{Institut d'Astrophysique de Paris, 98 bis Boulevard
Arago, 75014 Paris, France}
\author{E.J. Copeland}
\email[Electronic address: ]{e.j.copeland@sussex.ac.uk}
\affiliation{Centre for Theoretical Physics, University of Sussex, 
Brighton BN1 9QJ, United Kingdom}

\begin{abstract}

We discuss the onset of the non-linear regime in the context of unified 
dark matter models 
involving a generalised Chaplygin gas. We show that the transition from 
dark matter-like to dark energy-like behaviour will never be smooth. In 
some regions of space the transition will never take place while in others it 
may happen sooner or later than naively expected. As a result the linear 
theory used in previous studies may break down late in the matter dominated 
era even on large cosmological scales. We study the importance of this effect 
showing that its magnitude depends on the exact form of the equation of state 
in the low density regime. We expect that our results will be relevant for 
other unified dark matter scenarios particularly those where the quartessence 
candidate is a perfect fluid.

\end{abstract}
\date{24 June 2003}
\pacs{98.80.Es, 12.60.-i, 95.35.+d}
\keywords{Cosmology; Dark matter; Baryons}
\preprint{DAMTP-2003-55}
\maketitle

\section{Introduction}

There is growing evidence that the Universe we live in is (nearly) flat and 
presently undergoing an accelerating phase \cite{Perlmutter,Riess,Tonry,wmap}.
A `dark' energy component is thought to be
responsible for this acceleration and either a `standard' cosmological 
constant \cite{lambda}, a quintessence scalar field \cite{Wang}, or a 
k-essence field \cite{Picon} are 
usually put forth to account for it. However, all current explanations 
necessarily face some level of fine-tuning, which motivates a search for 
further alternatives (see for example \cite{solid,brane,Bagla}).

The Chaplygin gas provides one such interesting alternative. It bears the 
exotic equation of state
\begin{equation}
p=-\frac{C}{\rho^\alpha}\,,
\label{eqnstate}
\end{equation}
where $\rho$ is the density, $C$ is a positive constant and 
$0\leq\alpha\leq1$. In a homogeneous universe its main property 
(cosmologically speaking) is that of mimicking cold dark matter at 
early times but progressively evolving into a cosmological constant 
later on. If $\alpha=0$, the Chaplygin gas model is 
in fact identical to a familiar $\Lambda$CDM scenario where 
$\Omega^*_m \equiv 1-\overline A$ is the equivalent 
matter density (with $\overline A=C/\rho^{1+\alpha}_0$ where $\rho_0$ is 
the Chaplygin gas density at the present time).

This particular behaviour indicates that dark energy and dark matter could 
just be different manifestations of a unique perfect fluid known as 
`quartessence'. In this regard, the Chaplygin gas 
\cite{Kamen,Bilic} 
(or simple generalisations thereof \cite{Bento}) has 
attracted considerable attention as a quartessence prototype since 
a connection between string theory and the original Chaplygin gas 
has been proposed (see for example \cite{Hassaine} and references therein).

The simple picture described above has been taken for granted in all 
the work on this subject so far, namely when confronting models with
cosmological observations \cite{Fabris1,Fabris,Avelino,Dev,Gorini,Makler,Alcaniz,Finelli,Sandvik,Bean,Bento1,Beca,Amendola}. There is, however, a 
\textit{caveat} to it which may have 
important implications.
As we previously stated, the Chaplygin gas starts out behaving as CDM,
with linear perturbations growing in the usual way as long as this regime
holds. In linear theory, once the background evolution starts switching to a 
$\Lambda$-like behaviour, perturbations in the Chaplygin gas will become 
heavily damped while the growth of baryonic perturbations is slowed 
considerably.
This would all be fine, in the sense that assuming linear theory to hold, 
regions of parameter space have been 
shown to exist where structure formation proceeds in a way that is in 
agreement with observations. However, there is no 
\textit{a priori} guarantee that non-linear effects will not significantly 
alter these conclusions.

Specifically, in this letter we show that in a class of unified dark matter 
scenarios this transition from dark matter-like to dark energy-like 
behaviour will never be smooth. The reason is simple:
at some point, density perturbations on a given scale reach 
the non-linear regime and consequently a large fraction of the mass is then 
expected to be incorporated in collapsed objects whose 
evolution is effectively decoupled from the background.
In these regions the Chaplygin gas never evolves into the dark energy-like 
stages (which only happens at sufficiently low densities). 
On the other hand, in low density regions the absolute value of the 
(negative) pressure may be much larger than the average value thus having the 
opposite effect on the dynamics of the universe (favouring an earlier 
accelerating phase).

\begin{figure}
\includegraphics[width=3.5in]{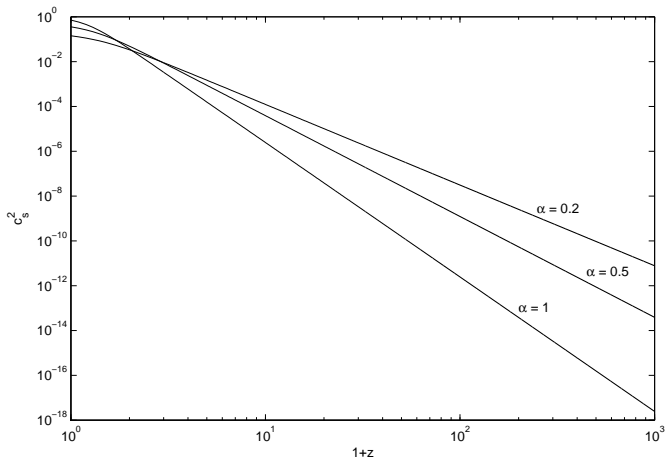}
\caption{\label{fig1} The evolution of the sound speed squared, $c_s^2$, 
of the Chaplygin gas as a function of redshift $z$ for $\alpha=0.2$, $0.5$ and 
$1$. We clearly see that at early times the Chaplygin gas closely resembles 
CDM.}
\end{figure}

It is the importance of this effect that we set out to quantify in this 
letter. While the analysis will be done for the case of the Chaplygin gas, 
our results will be relevant for any similar scenarios 
where the quartessence candidate is a perfect fluid. We note that 
concerns about the relevance of the non-linear evolutionary phase 
in a different context (condensation and metamorphosis cosmologies) have also 
been made in ref. \cite{Bassett}.

\section{The critical scales}

It is straightforward to show that in a homogeneous and isotropic 
universe the Chaplygin gas energy density evolves as 
\begin{equation}
\rho  = \rho _0 \left[ {\overline{A} +  (1 - \overline{A})
(1+z)^{3(1 + \alpha )} } \right]^{1/1 + \alpha},
\end{equation}
where $\overline{A}=C/\rho_0^{1+\alpha}$ and $\rho _0$ is the present 
density. In Fig. \ref{fig1} we plot the evolution of the sound speed squared 
$c_s^2=\alpha C /\rho^{1+\alpha}=-\alpha p/\rho$ as a function of the 
redshift $z$ taking $\overline{A}=0.71$ for three different values of 
$\alpha$ ($0.2$, $0.5$ and $1$). At early times the sound speed squared is 
\begin{equation}
c_s^2 \propto (1+z)^{-3(1+\alpha)},
\end{equation}
which determines how the inclination of the curves in Fig. \ref{fig1} depends  
on $\alpha$. We clearly see 
that deep in the matter era the equation of state of the Chaplygin gas 
closely resembles that of CDM (with $|p/\rho| \ll 1$).

Using a first order perturbative analysis, the evolution in Fourier space, of density perturbations in the Chaplygin gas, can be described by the following equation (see \cite{Sandvik} for details): 
\begin{eqnarray}
\delta_k''& + &[2+\xi -3(2w-c_s^2)]\delta_k'\\ \nonumber
&=& \left[\frac 32
(1-6c_s^2+8w-3w^2)-\left({k c_s\over a{\rm{H}} }\right)^2
\right]\delta_k\,.
\label{eq.pert}
\end{eqnarray}
which we rewrite as $\delta_k'' + A\delta_k' +B\delta_k = 0$. Here $a$ is 
the scale factor, $H$ is the Hubble parameter, $'\equiv d/d\ln a$, 
$\xi \equiv H'/H$ and $w \equiv p/\rho$.
This second order differential equation can easily be transformed into a non-autonomous dynamical system of the form ${\mbox{\boldmath $x$}'} = \mbox{\boldmath $A$} \, \mbox{\boldmath $x$}$ where $\mbox{\boldmath $x$}  \equiv (\delta, \delta')$ and $\mbox{\boldmath $A$}$ is the matrix of the coefficients, which are functions of both the wave number $k$ and the redshift $z$.

\begin{figure}
\includegraphics[width=3.5in]{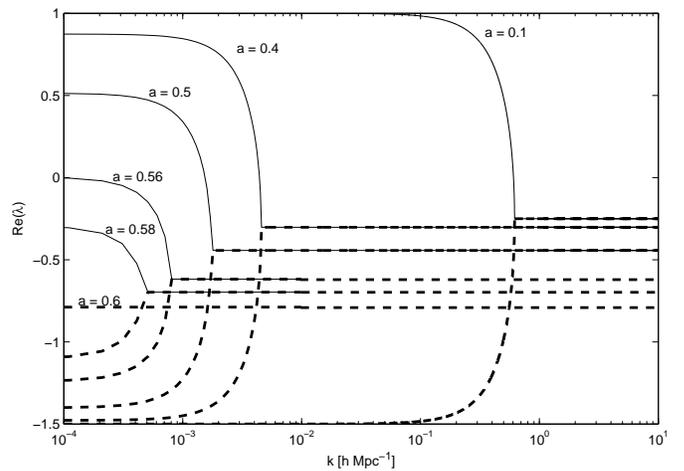}
\caption{\label{fig2}  The real part of the eigenvalues of the Chaplygin gas. The condition ${\rm{Re}}(\lambda(k_c, a))=0$ sets the critical scale for that instant. The solid line represents the first eigenvalue and the dashed, the second. Notice the bifurcation pattern.}
\end{figure}

We shall look for critical scales associated with the Chaplygin gas by calculating the eigenvalues $\lambda$ of $\mbox{\boldmath $A$} $ for fixed instants of time. These are the roots of the Cayley polynomial $\lambda^2+A\lambda+B=0$. Since the sign of the real part of the eigenvalue determines whether or not a mode grows or decays, the condition ${\rm{Re}}(\lambda(k_c, a))=0$ sets the critical scale for that $a$. Fig. 2 contains the real part of the two eigenvalues for several times. A bifurcation pattern emerges. At early times the Chaplygin gas acts as CDM and so all relevant scales are gravitationally unstable. As time goes by, this critical scale starts to diminish until there are no gravitational unstable scales. 

We note that in this linear analysis both $c_s$ and $w$ are evaluated to 
zeroth order. We will show that this is no longer a good approximation 
when the fluctuations in the Chaplygin gas component become large.

\section{Non-linear effects}

Having gained some intuition for the relevant length and time scales in
the problem, we now proceed with a more detailed analysis. Throughout the 
analysis we adopt priors in agreement with the WMAP first-year data 
release with a power-law  initial spectrum \cite{wmap}. The parameters are 
an equivalent matter density $\Omega_m^*=1 -\overline A + \Omega_b = 0.29$, 
a baryon density $\Omega_b=0.047$, an equivalent cosmological constant density 
$\Omega_\Lambda^* = 1-\Omega_m^* = 0.71$, a Hubble parameter 
$h=0.72$, a normalisation $\sigma_8=0.9$, and perturbation spectral index 
$n_{{\rm s}}=0.99$.

\begin{figure}
\includegraphics[width=3.5in]{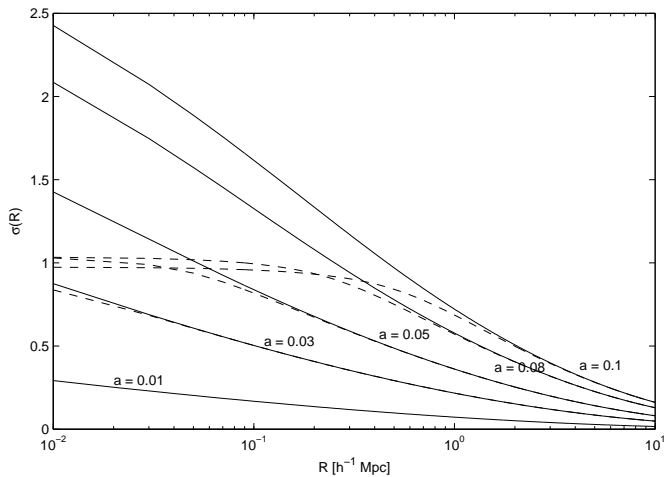}
\caption{\label{fig3} The linear evolution of the dispersion, $\sigma$, 
of the density 
fluctuations in the baryon (solid line) and Chaplygin gas (dashed lines) 
components as a function of $R$ and $a$ assuming $\alpha=1$. Note that at 
early times the baryon and Chaplygin gas fluctuations evolve in the same 
manner. Later on, pressure effects prevent the Chaplygin gas from collapsing 
further. However the baryon fluctuations can still 
keep growing (at a slower pace). 
}
\end{figure}

We start by determining how the linear density perturbations of the 
Chaplygin gas and baryon components evolve with redshift (the details of 
such an analysis are described in Sect. II of our earlier work \cite{Beca}).
We emphasise that the use of linear theory early in the matter era is 
actually a good approximation on large enough scales. The effects of 
the breakdown of linear theory will only be important 
(on large cosmological scales) at late times when a smooth transition 
from a dark matter to a dark energy dominated universe would naively be 
expected to take place.  

Hence, we use linear theory in order to compute the value of the 
dispersion of the density fluctuations in the baryon and Chaplygin gas 
components, $\sigma(R,a)$, as a function of $R$ and $a$. This is plotted in 
Fig. \ref{fig3} for the particular case of the original Chaplygin gas 
with $\alpha=1$. We see that since the Chaplygin gas behaves as matter at 
early stages (see Fig. \ref{fig1}), embedded perturbations will grow
proportionally to the scale factor, $a$, and in tune with those in the 
baryonic 
component. Therefore, both fluids evolve in the same way earlier on and have 
approximately the same value of $\sigma$ on all relevant scales. 
During this stage we have $\sigma \propto a = (1+z)^{-1}$. Later on, the 
pressure of the Chaplygin gas will have increased dramatically preventing 
further collapse of the Chaplygin gas component. However, the baryon 
fluctuations can still keep growing (at a slower pace). We also see in 
Fig. \ref{fig3} that the Chaplygin gas component becomes non-linear on small 
scales early in the matter era. It is clear that when this happens a 
significant fraction of the Chaplygin gas will have collapsed and decoupled 
from the background so that a transition from a dark matter-like to a dark 
energy-like stage (which necessarily requires lower densities) never 
happens in those regions. Using the Press-Schechter framework 
\cite{Press} we can show that for $\sigma=1$ the fraction of the equivalent 
mass that is already incorporated in collapsed objects is already close to 
$0.1$.

\begin{figure}
\includegraphics[width=3.5in]{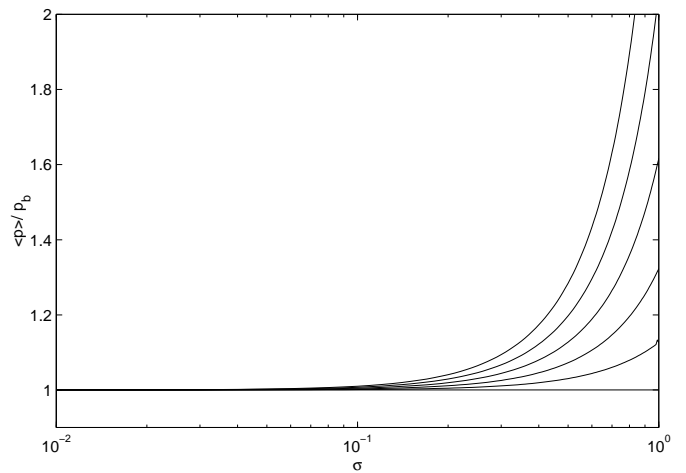}
\caption{\label{fig4} The ratio between the average value of $p$ and and its 
zeroth order background value, $\langle p \rangle /p_b$, 
as a function of the dispersion, $\sigma$, of the linear density fluctuations 
in the Chaplygin gas component for various values of $\alpha$ 
($1$, $0.8$, $0.6$, 
$0.4$, $0.2$ and $0$ from top to bottom). For $\alpha > 0$ we clearly see that 
$\langle p \rangle /p_b$ rapidly diverges from unity if $\sigma$ is large 
enough.
}
\end{figure}

In the non-linear regime an initial gaussian density field is better 
described by a lognormal one-point probability distribution function, 
${\cal P}(\delta)$, given by (see for example \cite{Sahni} and references 
therein)
\begin{equation}
{\cal P}(\delta)=\frac{(1+\delta)^{-1}}{\sqrt {2 \pi \ln(1+\sigma_{nl}^2)}}\exp\left(
-\frac{\ln^2\left((1+\delta){\sqrt {1+\sigma_{nl}^2}}\right)}{2 
\ln(1+\sigma_{nl}^2)}\right),
\label{lognormal}
\end{equation}
where $\sigma_{nl}^2=\exp(\sigma^2)-1$ and $\sigma$ is computed using linear 
theory. We use (\ref{lognormal}) in order to calculate the ratio between 
the average value of $p$ and its zeroth order background value 
($p_b=-C/\rho_b^\alpha$),
\begin{equation}
\langle p \rangle / p_b=\int_{-1}^\infty 
(1+\delta)^{-\alpha}{\cal P}(\delta) d \delta\,,
\label{epsilon}
\end{equation}
as a function of the dispersion of the density fluctuations in the 
Chaplygin gas component, $\sigma$. The results are displayed in Fig. 
\ref{fig4} 
for various values of $\alpha$ ($0$, $0.2$, $0.4$, $0.6$, $0.8$ and $1$). 
We see that in all but one case (for $\alpha=0$) the average value of the 
pressure 
strongly diverges from its zeroth order background value as soon as 
$\sigma$ becomes large enough. This means that the average values of both 
$w$ and $c_s^2$ will also move away from their zeroth order value 
(except in the $\alpha=0$ case) causing the breakdown of linear theory. 
The magnitude of this effect becomes 
more pronounced at late times when the negative pressure starts to become 
dynamically important on all scales.

We further note that in the $\alpha \neq 0$ case and for very small densities 
the sound speed, $c_s \propto \rho^{-(1+\alpha)/2}$, will become greater than 
the speed of light, $c=1$. Hence, 
we may expect that in the context of realistic models there will be a cut-off 
to this power law behaviour. One equation of state that resembles that of 
eqn. (\ref{eqnstate}) at high densities while switching off to 
a cosmological constant like equation of state at low densities is
\begin{equation}
p=-\frac{\Delta^{1+\beta} \rho}{(\rho+\Delta)^{1+\beta}}\,,
\label{cutoff}
\end{equation}
where $\Delta \ge 0$ is a low density cut-off and $0 \le \beta \le 1$. Note that 
in this case the absolute values of both $c^2_s$ and $w$ are always smaller 
than unity. Another 
interesting case is 
\begin{equation}
p=-\frac{C}{(\rho+\Delta)^{\beta}}\,,
\label{cutoff1}
\end{equation}
for which $c^2_s \le 1$ (assuming that $C \beta / \Delta^{\beta+1} \le 1$) 
but the absolute value of $w$ can grow arbitrarily large. 
As we have shown the importance of the non-linear effects discussed in this 
letter for cosmological predictions will strongly depend on the exact form 
of the equation of state in the low density regime.

\section{Conclusions}

We have shown that previous work on the Chaplygin gas has a 
\textit{caveat} which may have crucial implications for the predicted 
observational consequences of the model. The linear theory used in previous 
treatments breaks down at late times even on large cosmological scales 
except in the $\alpha=0$ case. This means that non-linear effects should 
be taken into account when confronting the model with cosmological 
observations. 

Although the present analysis has been done for 
the particular case of a Chaplygin gas we expect our results to be relevant  
for other unified dark matter scenarios particularly those where the 
quartessence candidate is a perfect fluid. 

\begin{acknowledgments}
C.M. is funded by FCT (Portugal), under grant FMRH/BPD/1600/2000.
Additional support came from FCT under contract CERN/POCTI/49507/2002.
\end{acknowledgments}

\bibliography{nonlinear}

\end{document}